# Observation of Pulsed $\gamma$–Rays Above 25 GeV from the Crab Pulsar with MAGIC

The MAGIC collaboration[1]

**One fundamental question about pulsars concerns the mechanism of their pulsed electromagnetic emission. Measuring the high-end region of a pulsar's spectrum would shed light on this question. By developing a new electronic trigger, we lowered the threshold of the Major Atmospheric gamma-ray Imaging Cherenkov (MAGIC) telescope to 25 GeV. In this configuration, we detected pulsed gamma-rays from the Crab pulsar that were greater than 25 GeV, revealing a relatively high cutoff energy in the phase-averaged spectrum. This indicates that the emission occurs far out in the magnetosphere, hence excluding the polar-cap scenario as a possible explanation of our measurement. The high cutoff energy also challenges the slot-gap scenario.**

It is generally accepted that the primary radiation mechanism in pulsar magnetospheres is synchrotron-curvature radiation. This occurs when relativistic electrons are trapped along the magnetic field lines in the extremely strong field of the pulsar. Secondary mechanisms include ordinary synchrotron and inverse Compton scattering. It is not known whether the emission of electromagnetic radiation takes place closer to the neutron star (NS) [the polar-cap scenario (1–3)] or farther out in the magnetosphere [the slot-gap (4–6) or outer-gap (7–9) scenario (Fig. 1)]. The high end of the gamma-ray spectrum differs substantially between the near and the far case. Moreover, current models of the slot gap (6) and the outer gap (8, 9) differ in their predicted gamma-ray spectra, even though both gaps extend over similar regions in the magnetosphere. Therefore, detection of gamma-rays above 10 GeV would allow one to discriminate between different pulsar emission models.

At gamma-ray energies (E) of ~1 GeV, some pulsars such as the Crab (PSR B0531+21) are among the brightest gamma-ray sources in the sky. The Energetic Gamma-ray Experiment Telescope (EGRET) detector, aboard the Compton gamma-ray Observatory (CGRO), measured the gamma-ray spectra of different pulsars only up to $E \approx 5$ GeV because of its small detector area (~0.1 $m^2$) and the steeply falling gamma-ray fluxes at higher energies. At $E > 60$ GeV, Cherenkov telescopes (10) are the most sensitive instruments because of their large detection areas of $\geq 10^4$ $m^2$. But, in spite of several attempts, no pulsar has yet been detected (11–16). This suggests a spectral cutoff; that is, that the pulsar's emission drops off sharply, between a few giga– electron volts and a few tens of GeV.

The Crab pulsar is one of the best candidates for studying such a cutoff. Its spectrum has been measured by EGRET (17) up to $E \approx 5$ GeV without a cutoff being seen. Earlier observations with the 17 m-diameter Major Atmospheric Gamma-ray Imaging Cherenkov (MAGIC) (18) telescope (Canary Island of La Palma, 2200 m above sea level) revealed a hint of pulsed emission at the 2.9 standard deviation ($\sigma$)

[1] The full list of authors and affiliations is presented at the end of this paper

level above 60 GeV (19, 20). To verify this result, we developed and installed a new trigger system that lowered the threshold of MAGIC from ~50 GeV to 25 GeV [supporting online material (SOM) text] (21).

We observed the Crab pulsar between October 2007 and February 2008, obtained 22.3 hours of good-quality data, and detected pulsed emission above 25 GeV. The pulsed signal (Fig. 2) has an overall significance of 6.4 σ with 8500±1330 signal events. Phase zero (ϕ= 0) is defined as the position of the main radio pulse (22). Our E > 25 GeV data show pronounced pulses at ϕ = 0 (main pulse, P1) and at ϕ = 0.3 to 0.4 (interpulse, P2). These pulses are coincident in phase with those measured by EGRET at E > 100 MeV and those coming from our own optical measurement. P1 and P2 have similar amplitudes at E = 25 GeV, in contrast to measurements at lower energies of E > 100 MeV, at which P1 is dominant. The present data show a small excess (3.4 σ) above 60 GeV for P2, which is consistent with our previous Crab observation (19, 20).

For the Crab pulsar, EGRET measured a power-law spectrum [$F(E) \propto E^{-\alpha}$ with $\alpha = 2.022 \pm 0.014$; F is the flux] in the energy range from E = 0.1 GeV to 5 GeV (17). At E = 25 GeV we measured a flux that was several times lower than a straightforward extrapolation of the EGRET spectrum, which would require a spectral cutoff somewhere between 5 and 25 GeV. Pulsar emission scenarios predict a generalized exponential shape for the cutoff that may be described as $F(E) = A\, E^{-\alpha} exp(-(E/E_0)^{\beta})$, where A is a normalized constant, $E_0$ is the cutoff energy, and $\beta$ measures the steepness of the cutoff. To determine the relevant parameters, we performed a joint fit to the Imaging Compton Telescope (COMPTEL) (≈1 to 30 MeV), EGRET (≈30 MeV to 10 GeV), and MAGIC (>25 GeV) data. For the conventional cases of $\beta$ = 1 (exponential) and $\beta$ = 2 (superexponential), we found $E_0 = 17.7 \pm 2.8_{stat} \pm 5.0_{syst}$ GeV (stat, statistical error; syst, systematic error) and $E_0 = 23.2 \pm 2.9_{stat} \pm 6.6_{syst}$ GeV, respectively. For $\beta$ left a free parameter, the best fit yields $E_0 = 20.4 \pm 3.9_{stat} \pm 7.4_{syst}$ GeV and $\beta$=1.2. The systematic error is dominated by a possible mismatch between the energy calibrations of EGRET and MAGIC (SOM text).

From a theoretical point of view, the spectral cutoff is explained as a combination of the maximum energies that electrons (e) can reach (because of the balance between acceleration and radiation losses) and the absorption of the emitted gamma-rays in the magnetosphere. Absorption is controlled by two mechanisms: (i) magnetic e+-e− pair production in the extremely strong field close to the pulsar surface and (ii) photon-photon e+-e− pair production in dense photon fields. If, for a young pulsar like the Crab with a magnetic field B ~ $10^{12}$ to $10^{13}$ G, emission occurs close to the NS surface [as in classical polar-cap models (1–3)], then magnetic pair-production attenuation provides a characteristic super-exponential cutoff at relatively low energies; that is, a few GeV at most (3). If, on the other hand, emission occurs farther out in the magnetosphere, at several stellar radii or close to the light cylinder [as in slot-gap (4–6) and outer-gap (7–9, 23) models], then absorption mainly arising from photon-photon collisions sets in at higher energies and produces a shallower cutoff (roughly exponential in shape). In either case, however, the measured $E_0$ could be intrinsic to the emitted spectrum and hence would only provide an upper limit to the absorption strength.

Equation 1 of (3) (a largely model-independent relation derived from simulations of gamma-ray absorption by magnetic-pair production in rotating magnetic dipoles) relates the pair-creation cutoff energy, Emax, with the location of the emission region $r/R_0$ ($R_0$ is the NS radius; r is the distance of the emission region from the center of the NS) for a NS with surface magnetic field $B_0$ and period P:

$$E_{\max} \approx 0.4 \cdot \sqrt{P \frac{r}{R_0}} \cdot \max\left\{1, \frac{0.1 B_{crit}}{B_0} \left(\frac{r}{R_0}\right)^3\right\} \text{ GeV.}$$

The appropriate values for the Crab pulsar are $B_0 = 8 \times 10^{12}$ G (8), natural constant $B_{crit} = 4.4 \times 10^{13}$ G (3), and P = 0.033 s ($B_{crit}$ is the critical field that marks the onset of quantum effects in a magnetized plasma). Using for $E_{max}$ the superexponential cutoff energy $E_0$=23.2±2.9$_{stat}$±6.6$_{syst}$ GeV, derived above for $\beta$=2 as appropriate for the polar cap scenario, one obtains $r/R_0 > 6.2 \pm 0.2_{stat} \pm 0.4_{syst}$: i.e., the emitting region is located well above the NS surface. This result, however, contradicts the basic tenet of the polar-cap scenario (1–3) that particle acceleration and radiation emission do occur very close to the pulsar surface. This inconsistency rules out the polar-cap scenario for the Crab pulsar.

Our results therefore favor an outer-gap or maybe also slot-gap scenario for the Crab pulsar. For example, using in Eq. 1 the value of $E_0$ that corresponds to $\beta = 1$ (approximately consistent with the outer-gap picture), a high-altitude emitting region is inferred, which is fully consistent with the assumed scenario. Specific recent outer-gap (8, 9) and slot-gap (6) predictions are compared with our data in Fig. 4. Although the former can provide emission of photons of energies as high as 25 GeV and hence explain our gamma-ray data, recent predictions of the slot-gap model cannot. Thus, current outer-gap models seem preferred in explaining our measurement.

Lastly, our present measurements reveal a trend of P2/P1 increasing with energy: It is <0.5 at 100 MeV, ≈1 at 25 GeV, and >1 at 60 GeV (Fig. 2). This trend provides valuable information for theoretical studies that will further constrain the location of the emission region in the Crab pulsar's magnetosphere [for example, (9)].

## References and Notes

1. M. A. Ruderman, P. G. Sutherland, *Astrophys. J.*, 196, 51-72 (1975).

2. J. K. Daugherty, A. K. Harding, *Astrophys. J.* 252, 337-347 (1982).

3. M. G. Baring, *Adv. Space Res.*, 33, 552-560 (2004).

4. J. Arons, E. T. Scharlemann, *Astrophys. J.* 231, 854-879 (1979).

5. A. G. Muslimov, A. K. Harding, *Astrophys. J.* 606, 1143-1153 (2004).

6. A. K. Harding, J. V. Stern, J. Dyks, F. Frackowiak, *Astrophys. J.* 680, 1378-1393 (2008).

7. K. S. Cheng, C. Ho, M. Ruderman, *Astrophys. J.* 300, 500-539 (1986).

8. K. Hirotani, *arXiv*:0809.1283, (2008).

9. A. P. S. Tang, J. Takata, J. Jia, K. S. Cheng, *Astrophys. J.* 676, 562-572 (2008).

10. Gamma-rays induce particle air showers in the atmosphere that emit Cherenkov light. The detection of this light allows measuring the energy and the direction of the incident gamma-ray.

11. P. Chadwick, et al., *Astroparticle Physics*, 9, Is. 2, 131-136 (1998).

12. P. G. Edwards et al., *Astronomy & Astrophysics*, 291, 468-472 (1994).

13. F. Aharonian et al., *Astrophys. J.,* 614, 897-913 (2004).

14. M. de Naurois et al., *Astrophys. J.,* 566, 343-357 (2002).

15. R. W. Lessard et al., *Astrophys. J.,* 531, 942-948 (2000).

16. F. Aharonian et al., *Astron. &Astrophys.*, 466, 543-554 (2007)

17. L. Kuiper, W. Hermsen et al., Astron. & Astrophys., 378, 918-935 (2001).

18. The MAGIC telescope website (http://wwwmagic.mppmu.mpg.de).

19. A. N. Otte, *PhD Thesis Technical University Munich 2007* (http://mediatum2.ub.tum.de/doc/620881/document.pdf).

20. J. Albert, *et al.*, *Astrophys. J.* 674, 1037-1055 (2008).

21. The threshold of a Cherenkov telescope is usually defined as the peak in the energy distribution of triggered gamma ray events for a gamma ray source with an $E^{-2.6}$ power law, photon energy spectrum.

22. A. G. Lyne, R. S. Pritchard, F. Smith, *MNRAS*, 265, 1003 (1993).

23. R. W. Romani, *Astrophys. J.*, 470, 469-478 (1996).

24. P. Goldreich, W. H. Julian, *Astrophys. J.* 157, 869-880 (1969).

25. Data provided by EGRET (ftp://legacy.gsfc.nasa.gov/compton/data/egret/).

26. F. Lucarelli, J. A. Barrio, P., Antoranz, et al., *Nucl. Instr. Meth. A,* 589, 415-424 (2008).

27. Private communication, A. K. Harding.

28. We thank the electronics division at the MPI, Munich, for their work in developing and producing the analogue sum trigger system, especially O. Reimann, R. Maier, S. Tran and T. Dettlaff. We also thank L. Stodolsky for comments. We acknowledge the Instituto de Astrofísica for providing all infrastructure on the Roque de los Muchachos in La Palma. The support of the German BMBF and MPG, the Italian INFN and INAF, the Swiss SNF and Spanish MCINN is acknowledged. This work was also supported by ETH Research Grant TH 34/043, by the Polish MNiSzW Grant N N203 390834, and by the YIP of the Helmholtz Gemeinschaft.

# List of Authors and Affiliations

E. Aliu[1], H. Anderhub[2], L. A. Antonelli[3], P. Antoranz[4], M. Backes[5], C. Baixeras[6], J. A. Barrio[4], H. Bartko[7], D. Bastieri[8], J. K. Becker[5], W. Bednarek[9], K. Berger[10], E. Bernardini[11], C. Bigongiari[8,30], A. Biland[2], R. K. Bock[7,8], G. Bonnoli[12,], P. Bordas[13], V. Bosch-Ramon[13], T. Bretz[10], I. Britvitch[2], M. Camara[4], E. Carmona[7], A. Chilingarian[14], S. Commichau[2], J. L. Contreras[4], J. Cortina[1], M. T. Costado[15,16], S. Covino[3], V. Curtef[5], F. Dazzi[8], A. De Angelis[17], E. De Cea del Pozo[18], R. de los Reyes[4], B. De Lotto[17], M. De Maria[17], F. De Sabata[17], C. Delgado Mendez[15], A. Dominguez[19], D. Dorner[10], M. Doro[8], D. Elsässer[10], M. Errando[1], M. Fagiolini[12], D. Ferenc[20], E. Fernandez[1], R. Firpo[1], M. V. Fonseca[4], L. Font[6], N. Galante[7], R. J. Garcia Lopez[15,16], M. Garczarczyk[7], M. Gaug[15], F. Goebel[7], D. Hadasch[5], M. Hayashida[7], A. Herrero[15,16], D. Höhne[10], J. Hose[7], C. C. Hsu[7], S. Huber[10], T. Jogler[7], D. Kranich[2], A. La Barbera[3], A. Laille[20], E. Leonardo[12], E. Lindfors[21], S. Lombardi[8], F. Longo[17], M. Lopez[8], E. Lorenz[2,7], P. Majumdar[7], G. Maneva[22], N. Mankuzhiyil[17], K. Mannheim[10], L. Maraschi[3], M. Mariotti[8], M. Martinez[1], D. Mazin[1], M. Meucci[12], M. Meyer[10], J. M. Miranda[4], R. Mirzoyan[7], M. Moles[19], A. Moralejo[1], D. Nieto[4], K. Nilsson[21], J. Ninkovic[7], N. Otte[23,7,29], I. Oya[4], R. Paoletti[12], J. M. Paredes[13], M. Pasanen[21], D. Pascoli[8], F. Pauss[2], R. G. Pegna[12], M. A. Perez-Torres[19], M. Persic,[25] L. Peruzzo[8], A. Piccioli[12], F. Prada[19], E. Prandini[8], N. Puchades[1], A. Raymers[14], W. Rhode[5], M. Ribó[13], J. Rico[26,1], M. Rissi[2], A. Robert[6], S. Rügamer[10], A. Saggion[8], T. Y. Saito[7], M. Salvati[3], M. Sanchez-Conde[19], P. Sartori[8], K. Satalecka[11], V. Scalzotto[8], V. Scapin[17], T. Schweizer[7], M. Shayduk[7], K. Shinozaki[7], S. N. Shore[24], N. Sidro[1], A. Sierpowska-Bartosik[18], A. Sillanpää[21], D. Sobczynska[9], F. Spanier[10], A. Stamerra[12], L. S. Stark[2], L. Takalo[21], F. Tavecchio[3], P. Temnikov[22], D. Tescaro[1], M. Teshima[7], M. Tluczykont[11], D. F. Torres[26,18], N. Turini[12], H. Vankov[22], A. Venturini[8], V. Vitale[17], R. M. Wagner[7], W. Wittek[7], V. Zabalza[13], F. Zandanel[19], R. Zanin[1], J. Zapatero[5],

(MAGIC collaboration)

O.C. de Jager[27], E. de Ona Wilhelmi[1,28]

Correspondence and request of materials to Thomas Schweizer <tschweiz@mppmu.mpg.de> , Nepomuk Otte <otte@mppmu.mpg.de>, Michael Rissi, <Michael.Rissi@phys.ethz.ch>, Maxim Shayduk <shayduk@mppmu.mpg.de>, Marcos Lopez Moya <marcos.lopezmoya@pd.infn.it>

1. IFAE, Edifici Cn., Campus UAB E-08193 Bellaterra, Spain, 2. ETH, Zürich, CH-8093 Switzerland, 3. INAF, I-00136 Rome, Italy, 4. Universidad Complutense E-28040 Madrid, Spain, 5.Technische Universität Dortmund, D-44221 Dortmund, Germany, 6. Universitat Autònoma de Barcelona, E-08193 Bellaterra, Spain, 7. Max-Planck-Institut für Physik, D-80805 München, Germany, 8. Università di Padova and INFN, I-35131 Padova, Italy, 9. University of Lodz, PL-90236 Lodz, Poland, 10. Universität Würzburg D-97074 Würzburg, Germany, 11. DESY Deutsches Elektr.-Synchrotron, D-15738 Zeuthen, Germany,12. Università di Siena and INFN Pisa, I-53100 Siena, Italy, 13. Universitat de Barcelona (ICC/IEEC), E-08028 Barcelona, Spain, 14. Yerevan Physics Institute, AM-375036 Yerevan, Armenia, 15. IAC, E-38200, La Laguna,Tenerife, Spain, 16. Depto. de Astrofisica, Universidad, E-38206 La Laguna, Tenerife, Spain,17. Università di Udine and INFN Trieste, I-33100

Udine, Italy, 18. IEEC-CSIC, E-08193 Bellaterra, Spain, 19. CSIC, E-18080 Granada, Spain, 20. Davis University of California, CA-95616-8677, USA, 21. Tuorla Observatory, Turku University, FI-21500 Piikkiö, Finland, 22. Inst. for Nucl. Research and Nucl. Energy, BG-1784 Sofia, Bulgaria, 23. Humboldt-Universität zu Berlin, D-12489 Berlin, Germany, 24. Università di Pisa and INFN Pisa, I-56126 Pisa, Italy, 25. INAF/Osservatorio Astronomico and INFN, I-34143 Trieste, Italy, 26. ICREA, E-08010 Barcelona, Spain. 27. Unit for Space Physics, Northwest University, Potchefstroom 2520, South Africa 28. Now at Astroparticule et Cosmologie, CNRS, Universite Paris, F-75205 Paris Cedex 13, 29. Now at Santa Cruz Institute for Particle Physics, University of California, Santa Cruz, CA 95064, USA, 30. Now at IFIC - Instituto de Física Corpuscular , CSIC-UVEG, E-46071 Valencia, Spain

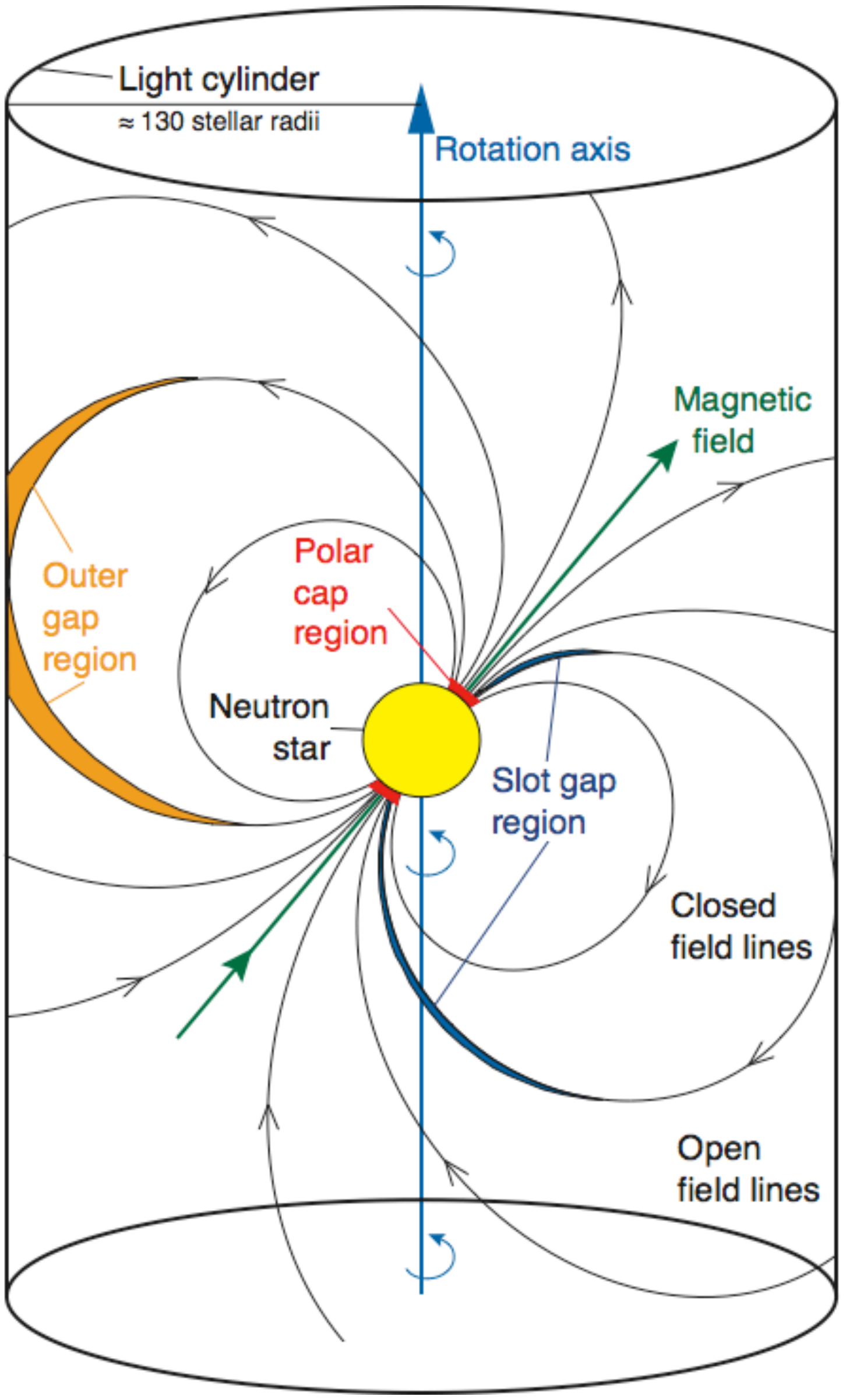

**Fig. 1: A sketch of the Crab pulsar's magnetosphere:** Electrons are trapped and accelerated along the magnetic field lines of the pulsar and emit electromagnetic radiation via the synchrotron-curvature mechanism. Vacuum gaps or vacuum regions occur at the polar cap (*1-3*), very close to the neutron star surface, in a thin layer extending for several stellar radii along the boundary of the closed magnetosphere, the so-called slot gap (*4-6*), and in the outer region (*7-9*) close to the light cylinder (outer gap). Vacuum gaps are filled with plasma, but its density is lower than the critical Goldreich-Julian density (*24*), where the magnetically induced electric field is saturated, and therefore electrons can be accelerated to very high energies. Absorption of high-energy γ-rays occurs by interaction with the magnetic field (magnetic pair production) as well as with the photon field (photon-photon pair production). The former dominates close to the surface of the neutron star where the magnetic field is strongest: it leads to a super-exponential cutoff at relatively low energies (few GeV). Photon-photon collisions prevail farther out in the magnetosphere close to light cylinder, where the magnetic field is lower, and lead to a roughly exponential cutoff at higher (>10 GeV) energies.

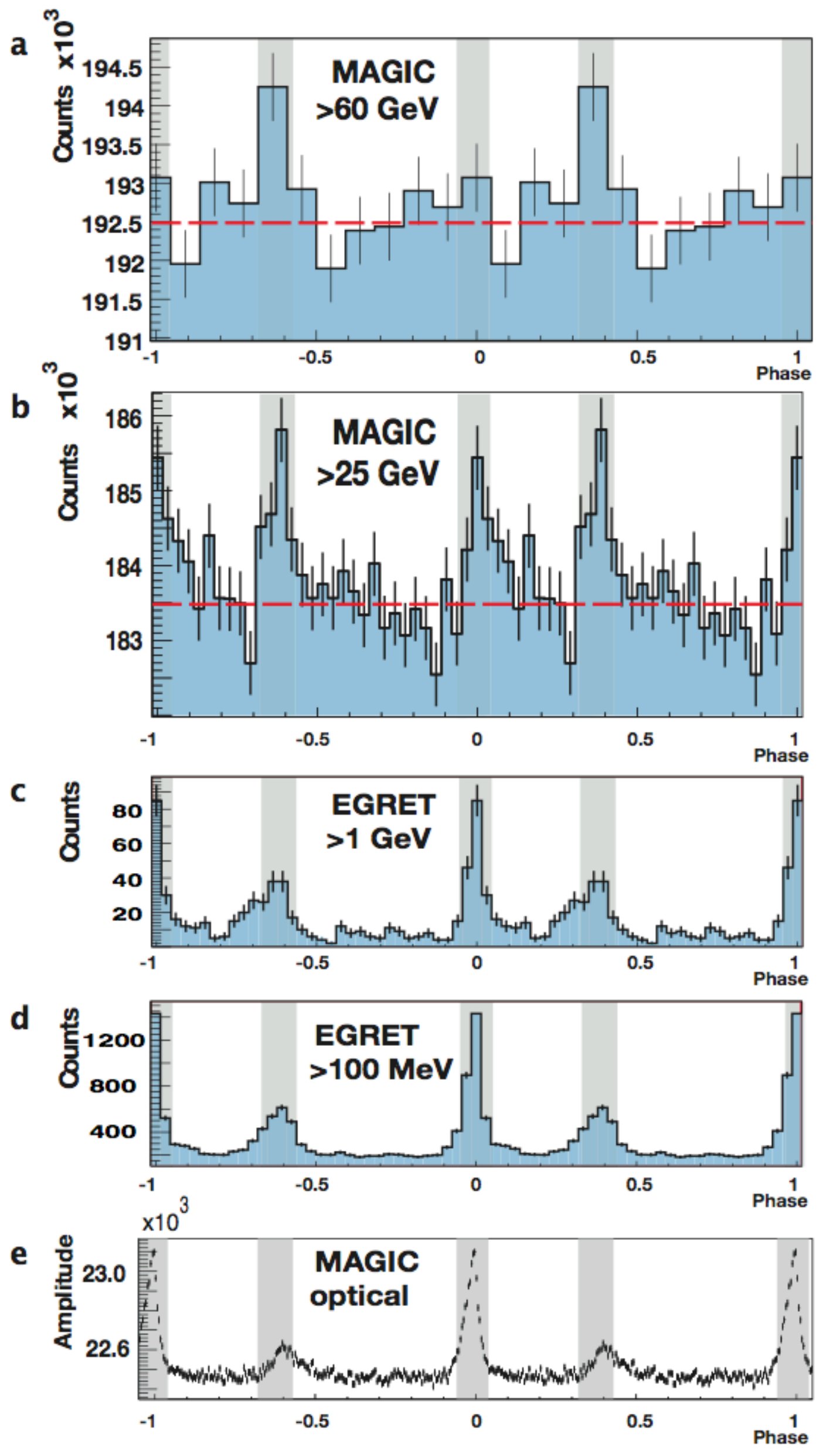

**Fig. 2: Pulsed emission in different energy bands**. The shaded areas show the signal regions for the main pulse (P1) and the inter pulse (P2). From top to bottom: (a) Evidence of an emission (3.4 σ) above 60 GeV for P2 measured by MAGIC; (b): Emission ≥ 25 GeV measured by MAGIC; (c) Emission ≥ 1 GeV measured by EGRET (*17*); (d) Emission ≥ 100 MeV measured by EGRET ( *25*); (e) Optical emission measured by MAGIC with the central pixel (*26*) of the camera. The optical signal has been recorded simultaneously with the γ–rays. P1 and P2 are in phase for all shown energies. The figure illustrates how the ratio of P2/P1 increases with energy in the panels (b)-(d). In search for pulsed emission, the arrival time of each event, after correcting for the solar system barycenter, was transformed into the phase of the rotational period of the neutron star. The significance of the γ-ray pulsation above 25 GeV was evaluated by a single-hypothesis test (described in the supporting material), where the γ-ray emission was assumed coming from the two fixed phase intervals (shaded regions): P1 (phase 0.94 to 0.04) and P2 (phase 0.32 to 0.43), as defined in (*19,20*). The signal results in 8500±1330 signal events (6.4σ).

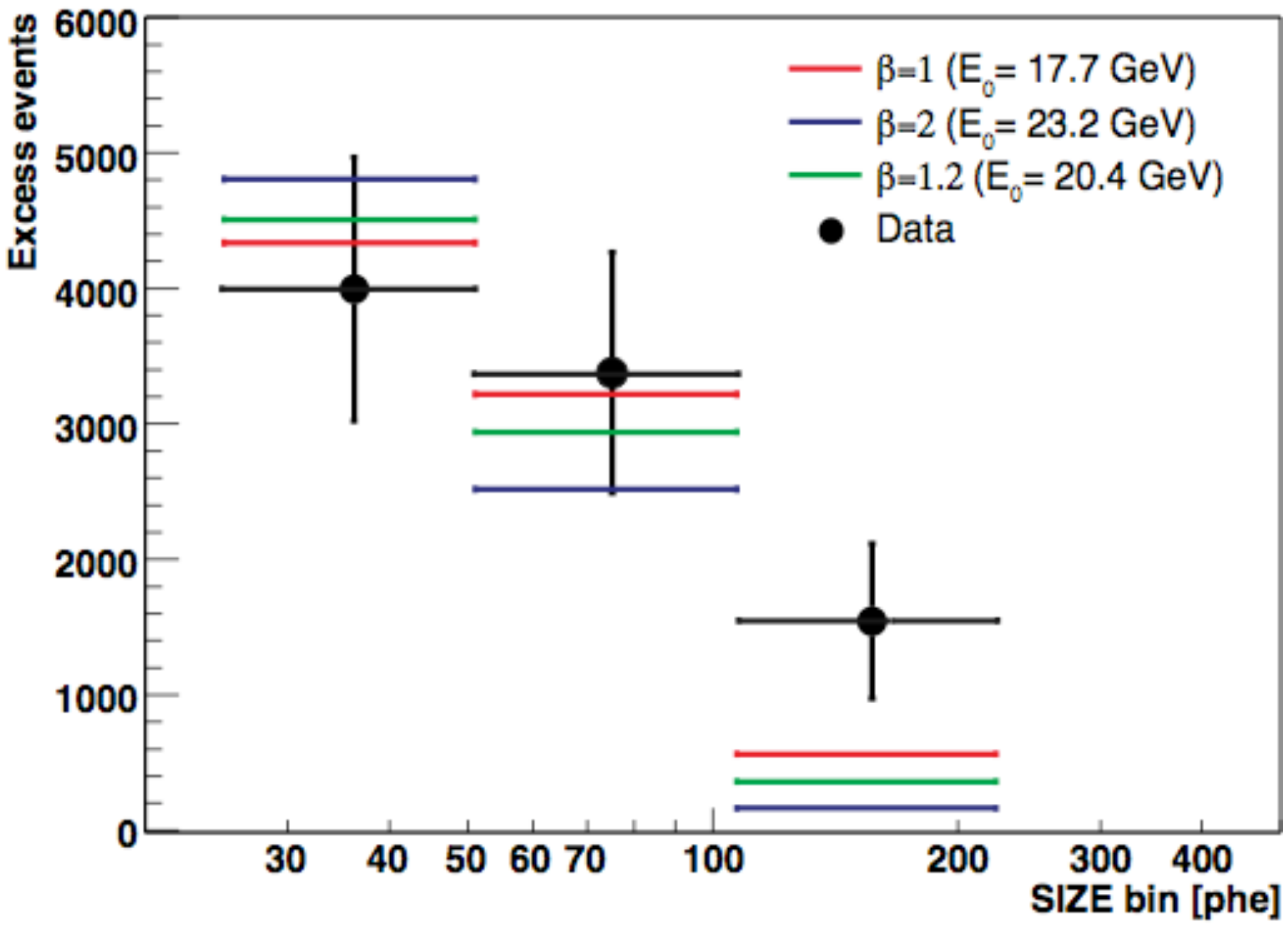

**Fig. 3. Model fits to the signal-event distribution**. Shown is the measured distribution of excess events in bins of SIZE integrated over P1 and P2. SIZE is a main image parameter that measures the total intensity of the Cherenkov flash in the camera in units of photoelectrons (phe). In this analysis, it was used as a rough estimate of the gamma-ray energy. To determine the cutoff energy, we folded the power law function with the generalized exponential shape function $F(E) = A\,E^{-\alpha}\,exp(-(E/E_0)^{\beta})$ with the MAGIC telescope's effective area to calculate the expected signal in each SIZE bin (forward unfolding). The expected signal was compared with the measured excess events by calculating a $\chi^2$ test. We obtained the best fit by minimizing the joint c2 between real data (from COMPTEL, EGRET, and MAGIC) and the generalized function. For the conventional cases of b = 1 (exponential) and b = 2 (superexponential), we found $E_0$=17.7±2.8$_{stat}$±5.0$_{syst}$ GeV and $E_0$=23.2±2.9$_{stat}$±6.6$_{syst}$ GeV, respectively. If instead we leave $\beta$ as a free parameter, the best fit yields $E_0$=20.4±3.9$_{stat}$±7.4$_{syst}$ GeV and $\beta$=1.2.

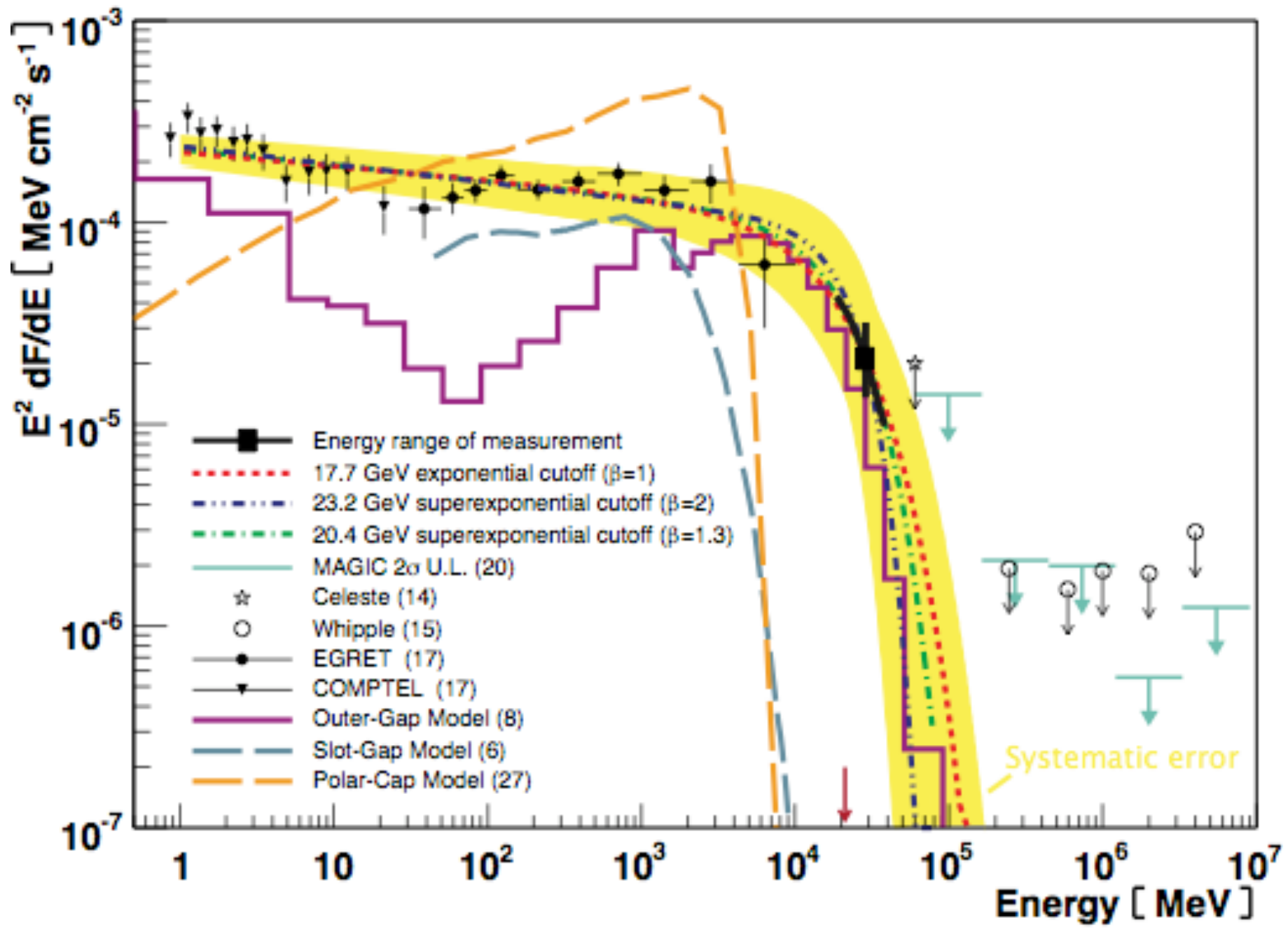

**Fig. 4: Crab pulsar spectral cutoff.** The black points and triangles on the left represent flux measurements from EGRET and COMPTEL (*17*). The arrows on the right denote upper limits from various previous experiments. We performed a joint fit of a function [$F(E) = A\,E^{-\alpha}\,exp(-(E/E_0)^{\beta})$] to the MAGIC, EGRET and COMPTEL data. The figure shows all three fitted functions for $\beta$=1 (red line) and $\beta$=2 (blue line) and the best fit $\beta$=1.2 (green line). The black line indicates the energy range, the flux and the statistical error of our measurement. The yellow band illustrates the joint systematic error of all three solutions. The measurement is compared with three current pulsar models, a polar cap model, a slot-gap model and an outer gap model. The sharp cut-off of the polar cap (*27*) model is due to magnetic pair production close to the surface of the neutron star. The slot gap model (*6*) does not reach the observed cut-off energy, while the outer gap (*8*) model can explain the high energy cut-off. The numbers in the parentheses refer to the list of references.